# Possible pairing mechanism switching driven by structural symmetry breaking in BiS$_2$-based layered superconductors

Aichi Yamashita[1], Hidetomo Usui[2], Kazuhisa Hoshi[1], Yosuke Goto[1], Kazuhiko Kuroki[3], Yoshikazu Mizuguchi[1]*

[1]*Department of Physics, Tokyo Metropolitan University, 1-1, Minami-osawa, Hachioji 192-0397, Japan.*

[2]*Department of Physics and Materials Science, Shimane University, 1060, Nishikawatsucho, Matsue 690-8504, Japan.*

[3]*Department of Physics, Osaka University, 1-1 Machikaneyama, Toyonaka, Osaka 560-0043, Japan.*

Corresponding author: Yoshikazu Mizuguchi (mizugu@tmu.ac.jp)

## Abstract

Investigation of isotope effects on superconducting transition temperature ($T_c$) is one of the useful methods to examine whether electron–phonon interaction is essential for pairing mechanisms. The layered BiCh$_2$-based (Ch: S, Se) superconductor family is a candidate for unconventional superconductors, because unconventional isotope effects have previously been observed in La(O,F)BiSSe and Bi$_4$O$_4$S$_3$. In this study, we investigated the isotope effects of $^{32}$S and $^{34}$S in the high-pressure phase of (Sr,La)FBiS$_2$, which has a monoclinic crystal structure and a higher $T_c$ of ~10 K under high pressures, and observed conventional-type isotope shifts in $T_c$. The conventional-type isotope effects in the monoclinic phase of (Sr,La)FBiS$_2$ are different from the unconventional isotope effects observed in La(O,F)BiSSe and Bi$_4$O$_4$S$_3$, which have a tetragonal structure. The obtained results suggest that the pairing mechanisms of BiCh$_2$-based superconductors could be switched by a structural-symmetry change in the superconducting layers induced by pressure effects.

# Introduction

In conventional superconductors, electron–phonon interactions are essential for the formation of Cooper pairs[1]. According to BCS (Bardeen-Cooper-Schrieffer) theory[1], the transition temperature ($T_c$) of a phonon-mediated superconductor is proportional to its phonon energy $\hbar\omega$, where $\hbar$ and $\omega$ are the Planck constant and the phonon frequency, respectively. Therefore, $T_c$ of conventional superconductors is sensitive to the phonon frequency, and modifications of the isotope mass ($M$) of the constituent elements, the so-called isotope effect, have been used to investigate the importance of electron–phonon interactions in the pairing of various superconductors. The isotope exponent $\alpha$ is defined by $T_c \sim M^{-\alpha}$, and $\alpha \sim 0.5$ is expected according to BCS theory[1]. For instance, $\alpha$ values close to 0.5 have been detected in $(Ba,K)BiO_3$ ($\alpha_O \sim 0.5$)[2], $MgB_2$ ($\alpha_B \sim 0.3$)[3], and borocarbides ($\alpha_B \sim 0.3$)[4]. In addition, the hydrides ($H_3S$ and $LaH_{10}$) high-$T_c$ superconductors also showed a conventional shift in $T_c$ with $\alpha_H = 0.3$–$0.5$ in isotope effect investigations[5,6]. In contrast, in superconductors with unconventional mechanisms, the isotope effect is not consistent with the BCS theory, and $\alpha$ values deviated from 0.5[7,8].

The target system of this study, layered $BiCh_2$-based (Ch: S, Se) superconductors, has been extensively studied since its discovery in 2012[9-11]. Because of its layered structure composed of alternate stacking of a superconducting layer and a blocking (insulating) layer, which resembles those of high-$T_c$ superconductors[12,13], many studies have been performed on material development and on pairing mechanisms[11]. Although non-doped (parent) $BiCh_2$-based compounds are semiconductors with a band gap, electron doping of the $BiCh_2$ layers makes the system metallic, and superconductivity is induced. An example of this is F substitution in $REOBiCh_2$ (RE: rare earth)[9-11]. In addition, the superconducting properties of $BiCh_2$-based systems are very sensitive to the effects of external (physical) and/or chemical pressures[14-17]. When external pressures are applied, the crystal structures of $REOBiCh_2$-based systems tend to distort into a monoclinic ($P2_1/m$) structure, and a higher-$T_c$ phase ($T_c > 10$ K) is induced[16]. Instead, by applying in-plane chemical pressure (shrinkage of the Bi-Ch conducting plane) via isovalent-element substitutions at the RE and/or Ch sites, a tetragonal ($P4/nmm$) phase is maintained, and bulk superconductivity is induced in the tetragonal phase. The emergence of bulk superconductivity due to chemical pressure effects can be explained by the suppression of local structural disorder, which is caused by the presence of Bi lone pair electrons[18-20].

Regarding the mechanisms of superconductivity in the $BiCh_2$-based family, the pairing mechanisms of the $BiCh_2$-based superconductor family are still controversial[19], owing to superconducting properties that are tunable by external and/or chemical pressure effects, which sometimes causes scattered results. Although earlier theoretical and experimental studies suggested conventional mechanisms with fully gapped s-wave pairing states[21-23], recent theoretical calculations of $T_c$ indicated that a $T_c$ of an order of several K to 10 K in $BiS_2$-based superconductors with a tetragonal structure cannot be explained within phonon-mediated models[24]. Furthermore, angle-resolved photoemission spectroscopy (ARPES) proposed unconventional pairing mechanisms owing to the observation of a highly anisotropic superconducting gap in $NdO_{0.71}F_{0.29}BiS_2$[25]. In addition, a study on the Se isotope effect with $^{76}Se$ and $^{80}Se$ in $LaO_{0.6}F_{0.4}BiSSe$ (Fig. 1f) indicated the possibility of unconventional (non-phonon) mechanisms with $\alpha_{Se}$ close to 0 (-0.04 < $\alpha_{Se}$ < 0.04)[26]. In addition, we have recently reported on an unconventional isotope effect with $^{32}S$ and $^{34}S$ in $Bi_4O_4S_3$ (-0.1 < $\alpha_S$ < 0.1) (Fig. 1g)[27]. These two superconductors have a tetragonal crystal structure and show a relatively low $T_c$ of 3.8 K for $LaO_{0.6}F_{0.4}BiSSe$ and 4.7 K for $Bi_4O_4S_3$. As mentioned above, the $BiS_2$-based superconductor has a high-pressure (high-$P$) phase, which exhibits a higher $T_c$ of over 10 K[16]. Therefore, this background encouraged us to plan an isotope effect study for a high-$P$ (monoclinic) phase with a higher $T_c$, in order to find a way to design new $BiCh_2$-based superconductors with a higher $T_c$ and to elucidate the mechanisms of superconductivity in the system.

Herein, we show experimental evidence of phonon-mediated superconductivity in a high-$P$ phase of $BiS_2$-based superconductors $(Sr,La)FBiS_2$. We have investigated the sulphur isotope effects ($^{32}S$ and $^{34}S$) on $T_c$ for a high-$P$ phase of $(Sr,La)FBiS_2$ with $T_c$ ~10 K[28-30]. Conventional shifts in $T_c$ between samples synthesised with $^{32}S$ and $^{34}S$ were observed, which suggests the importance of phonons for the pairing mechanisms in the compound. The conventional isotope effects in $(Sr,La)FBiS$, which has a monoclinic structure, are in contrast to the unconventional isotope effects observed in $La(O,F)BiSSe$ and $Bi_4O_4S_3$, which have a tetragonal structure[26,27]. Based on a combination of the discussion of previous and present isotope studies, we suggest that the structural difference between the tetragonal and monoclinic structures could be a switch of the pairing mechanisms in $BiCh_2$-based superconductors.

## Results
### Characterisation of isotope samples

In general, the shift in $T_c$ due to isotope effects is very small, even with $\alpha$ ~ 0.5 for low-$T_c$ superconductors. Therefore, examining the isotope effects with sets of samples with comparable superconducting properties is important to reach a reliable conclusion. However, precise control of the superconducting characteristics of $BiCh_2$-based compounds is the challenge of this study, because the $T_c$ of $BiCh_2$-based superconductors depends on the carrier concentration. From among the $BiCh_2$-based compounds, we selected the $Sr_{1-x}La_xFBiS_2$ system, because the carrier concentration in this system is essentially determined by the La concentration ($x$), and $x$ can easily be analysed by compositional analysis, such as energy dispersive X-ray spectroscopy (EDX). Here, we synthesised polycrystalline samples of $Sr_{1-x}La_xFBiS_2$ using $^{32}S$ and $^{34}S$ isotope chemicals for the investigation of sulphur isotope effects. We confirmed that the structural characteristics (particularly lattice constants) of the examined samples are comparable on the basis of powder X-ray diffraction (XRD) analyses (Figs. 1a and 1b). Detailed Rietveld analysis results are summarised in the Supplementary file. Although small impurity peaks of Bi and $LaF_3$ were observed, the lattice constants for the examined samples were comparable as shown in Fig. 1b. The La concentration ($x$) analysed by EDX was $x$ = 0.36~0.38, which is plotted in Fig. 1c. Among these samples, the carrier concentrations of samples #32-2, #34-1, and #34-2 were comparable, and that of #32-1 was slightly higher, where the sample labels indicate isotope mass (32 or 34) and batch number (1 or 2).

**Magnetisation measurements under high pressure**

As reported in a recent pressure study[30], (Sr,La)FBiS$_2$ shows a dramatic increase in $T_c$ from ~3 K for the low-pressure (low-$P$) phase to ~10 K for the high-$P$ phase on application of external pressure of about 1 GPa. The crystal structure of the high-$P$ phase can be regarded as monoclinic, whereas that for the low-$P$ phase is tetragonal, as shown in Figs. 1d and 1e, which is similar to the structural evolution of $LaO_{0.5}F_{0.5}BiS_2$ under pressures[16,30]. Figures 2a-2d show the temperature dependences of magnetisation measured at 10 Oe after zero-field cooling (ZFC). All samples of #32-1, #32-2, #34-1, and #34-2 show the transition from a low-$P$ phase to a high-$P$ phase, as plotted in Fig. 2e. Notably, in the high-$P$ phase after the $T_c$ jump, $T_c$ does not change by an increase in applied pressure below 1.4 GPa. A similar behaviour was reported for EuFBiS$_2$; the pressure dependence of $T_c$ of EuFBiS$_2$ showed a plateau under pressures above the critical pressure[31]. The appearance of the $T_c$ plateau would be related to the structural characteristics of BiS$_2$-based superconductors composed of fluoride-type blocking layers. This trend enabled us to examine the S isotope effect for the high-$P$ phase of the samples. Figure 3a shows selected data of the temperature dependence of magnetisation for high-$P$ phases of #32-1, #32-2, #34-1, and #34-2. Zoomed plots near the onset temperature of the superconducting transition ($T_c$) are shown in Fig. 3b. To estimate $T_c$, the temperature differential of magnetisation ($dM/dT$) was calculated and plotted as a function of temperature (Figs. 3c-3f). $T_c$ was estimated to be the temperature at which linear fitting lines for just below the transition temperature within a range of 0.5 K, as indicated by the red lines in those figures. The estimated $T_c$ are 10.42, 10.16, 9.94, and 9.73 K for the high-$P$ phases of #32-1, #32-2, #34-1, and #34-2, respectively (see Table I for the error). The highest $T_c$ was observed for #32-1 with a higher La concentration (electron doping amount). For the two samples with $^{34}$S, $x$ for #34-1 is slightly higher than $x$ for #34-2, while the difference is within the error bars shown in Fig. 1c. The difference in $T_c$, however, can be seen in Figs. 3e and 3f. The trend that a higher $T_c$ is observed for a sample with higher $x$ is consistent with the trend seen for #32-1 and #32-2. Although the $T_c$ is sensitive to the La concentration, we can reach a conclusion by comparing the $T_c$ based on the analysed La concentrations. When comparing the $T_c$ between #32-2 and #34-1, a different trend was observed; the $T_c$ estimated for #32-2 was higher than that of #34-1 with $x$ slightly higher than $x$ for #32-2. This fact implies that the isotope effect in the high-$P$ phase of $Sr_{1-x}La_xFBiS_2$ is conventionally present.

As La concentrations for #32-2 and #34-2 are very close, estimation of their $\alpha_S$ may be essential, which gives $\alpha_S \sim 0.7$. This value is slightly larger than the conventional $\alpha = 0.5$ expected from BCS theory, but it suggests the importance of phonon-mediated pairing in the high-$P$ phase of $(Sr,La)FBiS_2$. There are uncertainties in the determination of the essential $\alpha_S$ for the high-$P$ phase of $(Sr,La)FBiS_2$ because $T_c$ depends on the carrier concentration in this system, and the expected difference in $T_c$ between samples with $^{32}S$ and $^{34}S$ is not large. However, with the results shown here and the systematic analyses of $\alpha_S$, we can reach the conclusion that phonons are essential for the superconductivity pairing mechanisms in the high-$P$ phase of $(Sr,La)FBiS_2$. This is in contrast to the unconventional isotope effects observed in $La(O,F)BiSSe$[26] and $Bi_4O_4S_3$[27]. We discuss the possible differences in the structural and electronic characteristics of $(Sr,La)FBiS_2$ ($\alpha_S$ ~0.7) and $La(O,F)BiSSe$ ($-0.04 < \alpha_{Se} < 0.04$)[26] in the following section.

## Discussion

As summarised in Fig. 1, isotope effect suggesting the importance of phonon was observed for the high-$P$ phase of $(Sr,La)FBiS_2$, whereas unconventional isotope effects were observed in $La(O,F)BiSSe$ and $Bi_4O_4S_3$[26,27]. Although there are some possible factors, which could affect isotope effect, other than pairing states[32], we consider that the observed difference in isotope effect is essentially caused by the different pairing states between those systems. The reason for proposing the scenario is the recent observation of nematic superconductivity in $La(O,F)BiSSe$[33,34]; nematic superconductivity has been observed in unconventional superconductors like Fe-based and $Bi_2Se_3$-based superconductors[35,36]. Since nematic superconductivity emerges in both $LaO_{0.9}F_{0.1}BiSSe$ (tetragonal) and $LaO_{0.5}F_{0.5}BiSSe$ (tetragonal) with different carrier concentrations but with comparable structures of the BiSSe conducting layer, unconventional pairing states would commonly present in tetragonal BiCh$_2$-based superconductors with a tetragonal symmetry without structural distortion or local disorder. In contrast, nematic superconductivity was not observed in $Nd(O,F)BiS_2$[37], which is also tetragonal but has larger local structural disorder than $La(O,F)BiSSe$[11,17,19]. These facts suggest the importance of structural symmetry in the conducting layers and would support our scenario suggested in this article. Here, we discuss the possible differences in electronic states and pairing states between tetragonal and monoclinic phases.

The high-$P$ phase of (Sr,La)FBiS$_2$ has a monoclinic structure and a distorted in-plane structure in the BiS$_2$ layers[30]. In contrast, La(O,F)BiSSe and Bi$_4$O$_4$S$_3$ have tetragonal structures, in which the square Bi-Ch network forms a superconducting plane. Although the low-$P$ phase of (Sr,La)FBiS$_2$ is tetragonal, same as for La(O,F)BiSSe and Bi$_4$O$_4$S$_3$, bulk superconductivity is not observed at ambient pressure because of insufficient in-plane chemical pressure[17-19,30]. In the low pressure range, bulk superconductivity is induced, but the determination of $T_c$ is difficult because there are two possible superconducting transitions of the manometer (Pb) and the high-$P$ phase. Based on the isotope effects in the high-$P$ phase of (Sr,La)FBiS$_2$, La(O,F)BiSSe, and Bi$_4$O$_4$S$_3$, we suggest that structural symmetry breaking in the superconducting BiCh$_2$ layer is an essential factor in the switching of the isotope effect from unconventional to conventional.

We calculated the band structures of (Sr,La)FBiS$_2$ and La(O,F)BiSSe (see Supplementary file). Note that the calculated results for (Sr,La)FBiS$_2$ are based on the tetragonal structure of the low-$P$ phase, because structural parameters for the high-$P$ phase have not been experimentally obtained for (Sr,La)FBiS$_2$, and the structural relaxation was not successful for the high-$P$ phase in this work. One can determine that the shape of the Fermi surface is similar between (Sr,La)FBiS$_2$ and La(O,F)BiSSe, because the expected carrier doping amount is comparable. Therefore, we consider that the different isotope effects were due to the modifications of electronic and/or phonon characteristics induced by structural symmetry breaking in the monoclinic phase. According to previous theoretical calculations for the tetragonal and monoclinic phases of La(O,F)BiS$_2$, band splitting results from a structural transition from tetragonal (low-$P$ phase) to monoclinic (high-$P$ phase)[38]. In addition, the impact of interlayer coupling between two BiS$_2$ layers, caused by the structural symmetry breaking, on the electronic states was suggested as a possibility. The switching of isotope effects between the tetragonal and monoclinic phases may be linked to the formation of the Bi–Bi bonding in the high-$P$ phase in the present system. Let us remind that the theoretical study on the calculation of $T_c$ for LaO$_{0.5}$F$_{0.5}$BiS$_2$ by Morice et al.[24] was performed on a tetragonal unit cell. Their conclusion is consistent with the unconventional isotope effects observed in tetragonal La(O,F)BiSSe and Bi$_4$O$_4$S$_3$. If the same calculation of $T_c$ could be performed on a monoclinic unit cell, a $T_c$ of 10 K may be reproduced. For that, high-resolution structural analyses of the high-$P$ phase of (Sr,La)FBiS$_2$ are needed.

In conclusion, we synthesised (Sr,La)FBiS$_2$ polycrystalline samples with $^{32}$S and $^{34}$S isotope chemicals. With magnetisation measurements under high pressure, we investigated the sulphur isotope effects on $T_c$ for a high-$P$ phase of (Sr,La)FBiS$_2$. As a conventional shift in $T_c$ was observed, we suggested the importance of phonons for the pairing mechanisms for the high-$P$ phase. Based on comparisons with isotope effects in La(O,F)BiSSe and Bi$_4$O$_4$S$_3$, in which unconventional isotope effects have been observed, we suggest that structural symmetry breaking from tetragonal to monoclinic is a key factor for the switch of the isotope effects in the BiCh$_2$-based superconductor family.

**Methods**

Polycrystalline samples of (Sr,La)FBiS$_2$ were prepared by a solid-state reaction method in an evacuated quartz tube. Powders of La (99.9%), SrF$_2$ (99%), and Bi (99.999%) were mixed with powders of $^{32}$S (ISOFLEX: 99.99%) or $^{34}$S (ISOFLEX: 99.26%) with a nominal composition of Sr$_{0.5}$La$_{0.5}$FBiS$_2$ in an Ar-filled glove box. The mixed powder was pelletised, and sintered in an evacuated quartz tube at 700 ºC for 20 h, followed by furnace cooling to room temperature. The obtained compounds were thoroughly mixed, ground, and sintered under the same conditions as the first sintering. Except for the starting materials, the synthesis method was the same as our recent study on (Sr,La)FBiS$_2$[30].

The phase purity and crystal structure of the (Sr,La)FBiS$_2$ samples were examined by laboratory X-ray diffraction (XRD) by the $\theta$-$2\theta$ method with Cu-K$\alpha$1 radiation on a SmartLab (RIGAKU) diffractometer. The schematic images of crystal structures were drawn by VESTA[39] using structural data refined by Rietveld refinement using RIETAN-FP[40]. Through the XRD analyses, Bi and LaF$_3$ impurity phases were detected. The actual compositions of the examined samples were analysed using energy dispersive X-ray spectroscopy (EDX) on a TM-3030 instrument (Hitachi). The average value of $x_{EDX}$ was calculated using the data obtained for five points on the sample surface. Standard deviation was estimated and shown in Table I. Through the XRD analyses, small spots with La-rich compositions were found. The impurity phase will be LaF$_3$, since a LaF$_3$ phase was found in XRD.

The temperature dependence of the magnetisation at ambient pressure and under high pressures was measured using a superconducting quantum interference device (SQUID) on MPMS-3 (Quantum Design) after zero-field cooling (ZFC). Hydrostatic pressures were generated by the MPMS high-pressure capsule cell made of nonmagnetic Cu-Be, as described in our recent high pressure study on (Sr,RE)FBiS$_2$[30]. The sample was immersed in a pressure transmitting medium (Daphene 7373) and covered with a Teflon cell. The pressure at low temperature was calibrated based on the superconducting transition temperature of the Pb manometer. The magnetisation data shown in this paper contains background magnetisation. For sample #32-1, the maximum pressure was lower than that for other samples, which is due to the setup of high-pressure cell with a shorter piston stroke.

## Data availability
All relevant data are available from the corresponding authors upon reasonable request.

## Acknowledgements
The authors thank R. Jha and O. Miura for their assistance with the experiments. This work was partly supported by JSPS KAKENHI (Grant Nos. 18KK0076, 15H05886, and 19K15291) and the Advanced Research Program under the Human Resources Funds of Tokyo (Grant Number: H31-1).

## Author contributions
A.Y. and Y.M. led the project. A.Y., K. H., Y.G., and Y.M. explored a phase suitable for the investigation of the sulphur isotope effects. A.Y. synthesised the samples. A.Y. and Y.M. characterised the samples using XRD and EDX. A.Y. and Y.G. performed the magnetisation measurements under pressure. Theoretical calculations were carried out by H.U. and K.K. The manuscript was written by A.Y, H.U., and Y.M. with input from all co-authors.

## Competing Interests
The authors declare no competing interests.

## References

1. Bardeen, J., Cooper, L. N., & Schrieffer, J. R. Theory of superconductivity. *Phys. Rev.* 108, 1175–1204 (1957).
2. Hinks, D. G. et al. The oxygen isotope effect in $Ba_{0.625}K_{0.375}BiO_3$. *Nature* 335, 419–421 (1988).
3. Bud'ko, S. L. et al. Boron isotope effect in superconducting $MgB_2$. *Phys. Rev. Lett.* 86, 1877–1880 (2001).
4. Lawrie, D. D. & Franck, J. P. Boron isotope effect in Ni and Pd based borocarbide superconductors. *Physica C* 245, 159–163 (1995).
5. Drozdov, A. P. et al. Conventional superconductivity at 203 kelvin at high pressures in the sulfur hydride system. *Nature* 525, 73–76 (2015).
6. Drozdov, A. P. et al. Superconductivity at 250 K in lanthanum hydride under high pressures. *Nature* 569, 528–531(2019).
7. Tsuei, C. C. et al. Anomalous isotope effect and Van Hove singularity in superconducting Cu

oxides. *Phys. Rev. Lett.* 65, 2724–2727 (1990).

8. Shirage, P. M. et al. Inverse iron isotope effect on the transition temperature of the (Ba,K)Fe$_2$As$_2$ superconductor. *Phys. Rev. Lett.* 103, 257003(1–4) (2009).
9. Mizuguchi, Y. et al. BiS$_2$-based layered superconductor Bi$_4$O$_4$S$_3$. *Phys. Rev. B* 86, 220510(1–5) (2012).
10. Mizuguchi, Y. et al. Superconductivity in novel BiS$_2$-based layered superconductor LaO$_{1-x}$F$_x$BiS$_2$. *J. Phys. Soc. Jpn.* 81, 114725(1–5) (2012).
11. Mizuguchi, Y. Material development and physical properties of BiS$_2$-based layered compounds. *J. Phys. Soc. Jpn.* 88, 041001(1-19) (2019).
12. Bednorz, J. G. & Müller, K. A. Possible high $T_c$ superconductivity in the Ba−La−Cu−O system. *Z. Phys. B* Condensed Matter 64, 189–193 (1986).
13. Kamihara, Y. et al. Iron-based layered superconductor La[O$_{1-x}$F$_x$]FeAs (x = 0.05− 0.12) with $T_c$ = 26 K. *J. Am. Chem. Soc.* 130, 3296–3297 (2008).
14. Jha, R. & Awana, V. P. S. Effect of hydrostatic pressure on BiS$_2$-based layered superconductors: a review. *Nov. Supercond. Mater.* 2, 16–31 (2016).
15. Wolowiec, C. T. et al. Enhancement of superconductivity near the pressure-induced semiconductor-metal transition in BiS$_2$-based superconductors LnO$_{0.5}$F$_{0.5}$BiS$_2$ (Ln = La, Ce, Pr, Nd). *J. Phys.: Condens. Matter* 25, 422201(1-6) (2013).
16. Tomita, T. et al. Pressure-induced enhancement of superconductivity and structural transition in BiS$_2$-layered LaO$_{1-x}$F$_x$BiS$_2$. *J. Phys. Soc. Jpn.* 83, 063704(1-4) (2014).
17. Mizuguchi, Y. et al. In-plane chemical pressure essential for superconductivity in BiCh$_2$-based (Ch: S, Se) layered structure. *Sci. Rep.* 5, 14968(1–8) (2015).
18. Mizuguchi, Y. et al. The effect of RE substitution in layered REO$_{0.5}$F$_{0.5}$BiS$_2$: chemical pressure, local disorder and superconductivity. *Phys. Chem. Chem. Phys.* 17, 22090–22096 (2015).
19. Mizuguchi, Y. et al. Evolution of anisotropic displacement parameters and superconductivity with chemical pressure in BiS$_2$-based REO$_{0.5}$F$_{0.5}$BiS$_2$ (RE = La, Ce, Pr, and Nd). *J. Phys. Soc. Jpn.* 87, 023704(1-4) (2018).
20. Paris, E. et al. Role of the local structure in superconductivity of LaO$_{0.5}$F$_{0.5}$BiS$_{2-x}$Se$_x$ system. *J. Phys.: Condens. Matter* 29, 145603 (2017).
21. Suzuki, K. et al. Electronic structure and superconducting gap structure in BiS$_2$-based layered superconductors. *J. Phys. Soc. Jpn.* 88, 041008 (1-13) (2019).
22. Wu, S. F. et al. Raman scattering investigation of the electron–phonon coupling in superconducting Nd(O,F)BiS$_2$. *Phys. Rev. B* 90, 054519(1–5) (2014).
23. Yamashita, T. et al. Conventional s-wave superconductivity in BiS$_2$-based NdO$_{0.71}$F$_{0.29}$BiS$_2$ revealed by thermal transport measurements. *J. Phys. Soc. Jpn.* 85, 073707(1–4) (2016).
24. Morice, C. et al. Weak phonon-mediated pairing in BiS$_2$ superconductor from first principles.

*Phys. Rev. B* 95, 180505(1–6) (2017).

25. Ota, Y. et al. Unconventional superconductivity in the BiS$_2$-based layered superconductor NdO$_{0.71}$F$_{0.29}$BiS$_2$. *Phys. Rev. Lett.* 118, 167002(1–6) (2017).

26. Hoshi, K., Goto, Y., & Mizuguchi, Y. Selenium isotope effect in the layered bismuth chalcogenide superconductor LaO$_{0.6}$F$_{0.4}$Bi(S,Se)$_2$. *Phys. Rev. B* 97, 094509(1-5) (2018).

27. Jha, R. & Mizuguchi, Y. Unconventional isotope effect on transition temperature in BiS$_2$-based superconductor Bi$_4$O$_4$S$_3$. *Appl. Phys. Express* 13, 093001(1-5) (2020).

28. Lin, X. et al. Superconductivity induced by La doping in Sr$_{1-x}$La$_x$FBiS$_2$. *Phys. Rev. B* 87, 020504(1-4) (2013).

29. Jha, R., Tiwari, B., & Awana, V. P. S. Impact of Hydrostatic Pressure on Superconductivity of Sr$_{0.5}$La$_{0.5}$FBiS$_2$, *J. Phys. Soc. Jpn.* 83, 063707(1-4) (2014).

30. Yamashita, A. et al. Evolution of two bulk-superconducting phases in Sr$_{0.5}$RE$_{0.5}$FBiS$_2$ (RE: La, Ce, Pr, Nd, Sm) by external hydrostatic pressure effect. *Sci. Rep.* 10, 12880(1-8) (2020).

31. Guo, C. Y. et al. Evidence for two distinct superconducting phases in EuBiS$_2$F under pressure. *Phys. Rev. B* 91, 214512(1-5) (2015).

32. Bill, A., Kresin, V. Z. & Wolf, S. A. The Isotope Effect in Superconductors. arXiv:cond-mat/9801222.

33. Hoshi, K., Kimata, M., Goto, Y., Matsuda, T. D. & Mizuguchi, Y. Two-Fold-Symmetric Magnetoresistance in Single Crystals of Tetragonal BiCh$_2$-Based Superconductor LaO$_{0.5}$F$_{0.5}$BiSSe. *J. Phys. Soc. Jpn.* 88, 033704(1-4) (2019).

34. Hoshi, K., Kimata, M., Goto, Y., Miura, A., Moriyoshi, C., Kuroiwa, Y., Nagao, M. & Mizuguchi, Y. Two-fold symmetry of in-plane magnetoresistance anisotropy in the superconducting states of BiCh$_2$-based LaO$_{0.9}$F$_{0.1}$BiSSe single crystal. *J. Phys. Commun.* 4, 095028(1-7) (2020).

35. Song, C. L., Wang, Y. L., Cheng, P., Jiang, Y. P., Li, W., Zhang, T., Li, Z., He, K., Wang, L. Jia, J. F., Hung, H. H., Wu, C., Ma, X., Chen, X. & Xue, Q. K. Direct Observation of Nodes and Twofold Symmetry in FeSe Superconductor. *Science* 332, 1410-1413 (2011).

36. Yonezawa, S. Nematic Superconductivity in Doped Bi$_2$Se$_3$ Topological Superconductors. *Condens. Matter* 4, 2(1-20) (2019).

37. Hoshi, K., Sudo, K., Goto, Y., Kimata, M. & Mizuguchi, Y. Investigation of in-plane anisotropy of c-axis magnetoresistance for BiCh$_2$-based layered superconductor NdO$_{0.7}$F$_{0.3}$BiS$_2$. arXiv:2007.03235.

38. Ochi, M., Akashi, R., & Kuroki, K. Strong bilayer coupling induced by the symmetry breaking in the monoclinic phase of BiS$_2$-based superconductors. *J. Phys. Soc. Jpn.* 85, 094705(1-8) (2016).

39. Momma, K. & Izumi, F., VESTA: a three-dimensional visualization system for electronic and

structural analysis. *J. Appl. Crystallogr.* 41, 653-658 (2008).

40. Izumi, F. & Momma, K. Three-dimensional visualization in powder diffraction. *Solid State Phenom.* 130, 15-20 (2007).

Table I. Sample label, $x_{EDX}$, and $T_c$ for the isotope samples of $Sr_{1-x}La_xFBiS_2$.

| Sample label | $x_{EDX}$ | $T_c$ (K) | Applied pressure (GPa) |
|---|---|---|---|
| #32-1 | 0.387(8) | 10.42(13) | 1.21 |
| #32-2 | 0.361(12) | 10.14(17) | 1.36 |
| #34-1 | 0.368(8) | 9.92(21) | 1.37 |
| #34-2 | 0.361(8) | 9.71(31) | 1.39 |

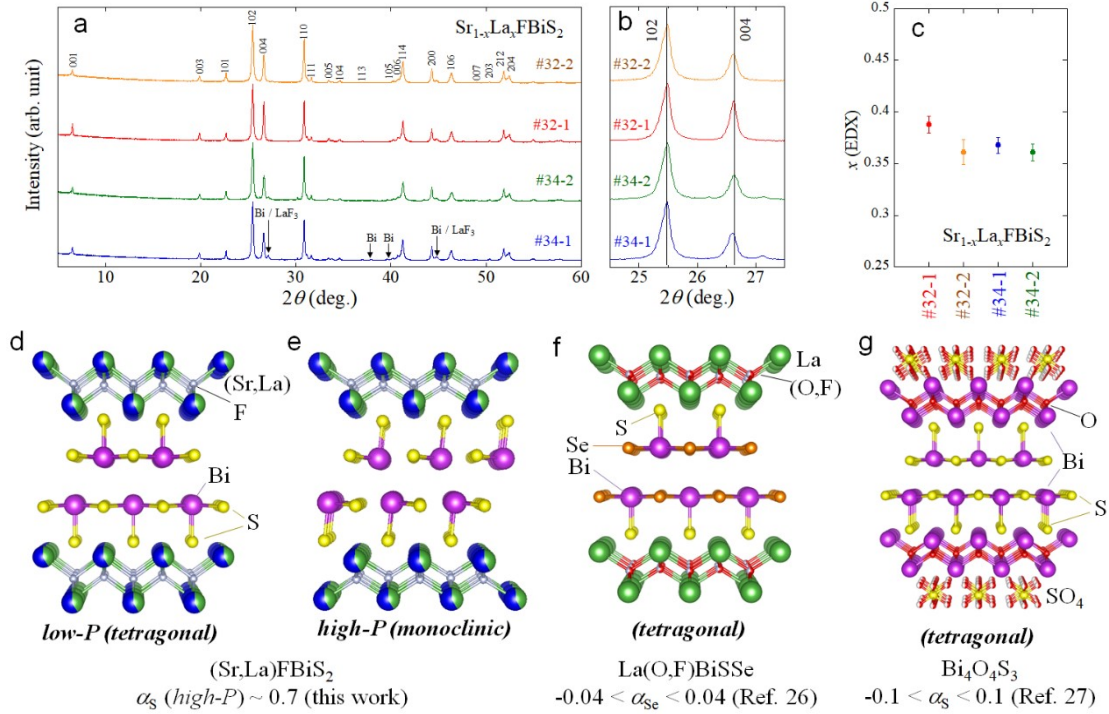

**Fig. 1. Structural and compositional data for $Sr_{1-x}La_xFBiS_2$ samples with different isotope mass for sulphur.** (a) Powder XRD patterns for #32-1, #32-2, #34-1, and #34-2. Numbers above the XRD pattern are Miller indices. Small amount of Bi and $LaF_3$ impurities were detected for #34-1 as indicated by arrows. (b) Zoomed XRD patterns near the 102 and 004 peaks. (c) La concentration ($x$) analysed by EDX. (d-g) Schematic images of crystal structure of the low-$P$ (tetragonal) phase and the high-$P$ phase (monoclinic) of $(Sr,La)FBiS_2$ and the tetragonal phase of $La(O,F)BiSSe$ and $Bi_4O_4S_3$. To emphasise the presence of quasi-one-dimensional network in the monoclinic phase (e), only the shorter Bi-S bonds were depicted. For comparison of the isotope effect exponent ($\alpha$) and the crystal structure, $\alpha_S$ for $(Sr,La)FBiS_2$, $\alpha_{Se}$ for $La(O,F)BiSSe$[26], and $\alpha_S$ for $Bi_4O_4S_3$[27] (a half unit cell) are shown.

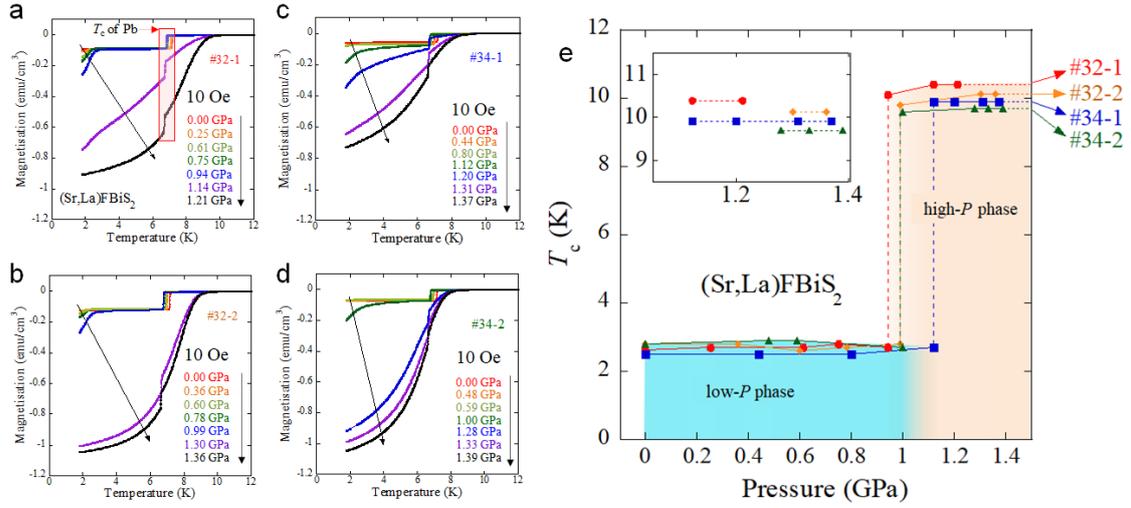

**Fig. 2. External pressure effects on the temperature dependence of magnetisation for isotope samples of $Sr_{1-x}La_xFBiS_2$.** (a-d) Temperature dependences of magnetisation for 32-1, #32-2, #34-1, and #34-2, respectively. Superconducting transitions at around 7 K are $T_c$ of the Pb manometer. (e) Pressure dependence of $T_c$. The inset shows the enlarged plot for the data of high-$P$ phases. Note that the $T_c$ for low-$P$ phases was roughly estimated because of superconducting signals mixed with those of the high-$P$ phase and the Pb manometer. See supplemental file for the estimation of the $T_c$ for the low-$P$ phase.

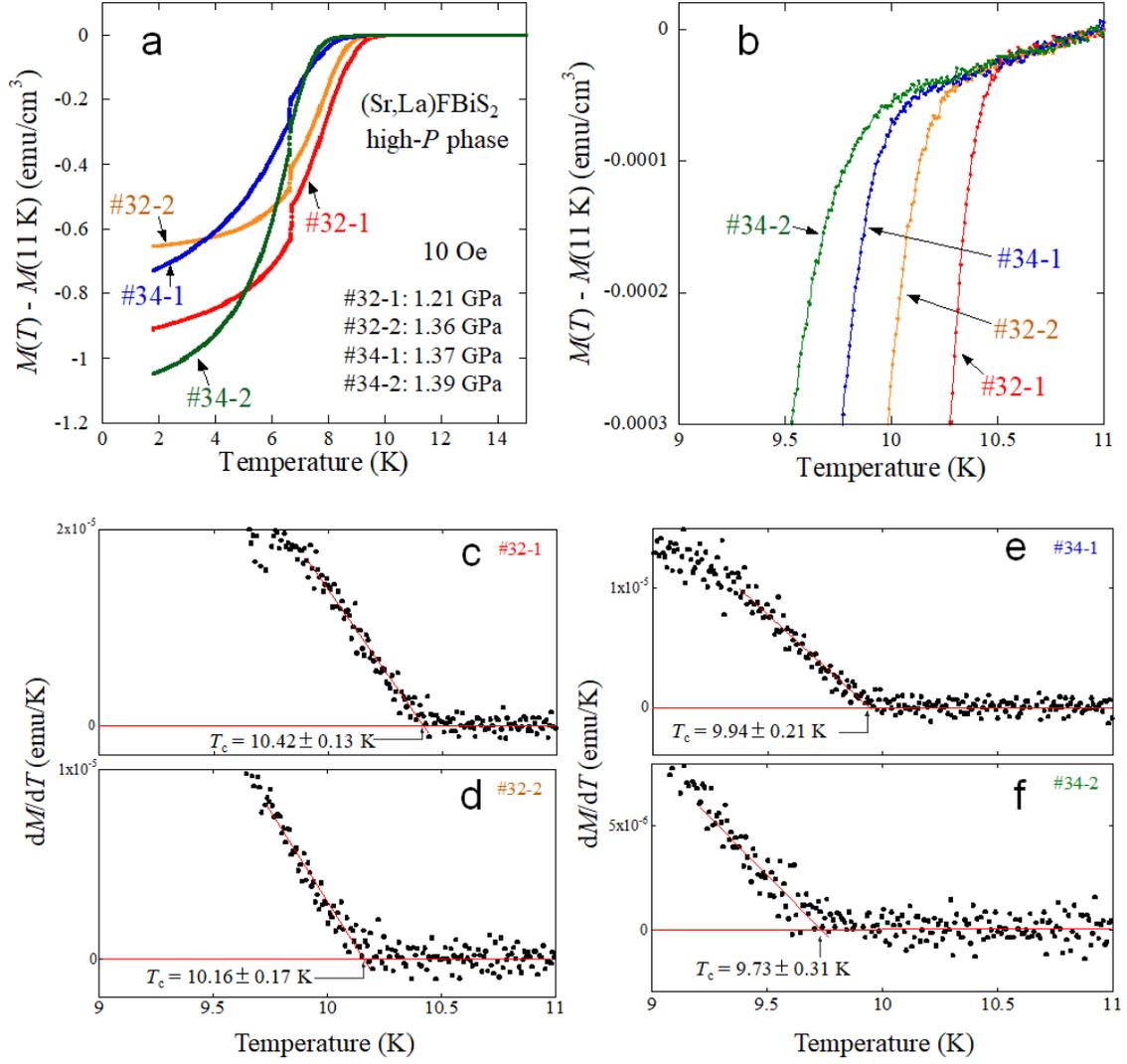

**Fig. 3. Estimation of $T_c^{onset}$ from data of the temperature dependences of magnetisation for isotope samples of $Sr_{1-x}La_xFBiS_2$.** (a) Temperature dependences of magnetisation for the high-$P$ phases of 32-1, #32-2, #34-1, and #34-2. (b) Zoomed figure of (a) near the $T_c$. (c-f) Temperature dependence of the temperature differential of magnetisation for #32-1, #32-2, #34-1, and #34-2. $T_c$ was estimated as the temperature at which linear fitting lines of just above and just below the onset of the transition cross as indicated by the red lines in the figures.

# Supplementary file

Table S1. Rietveld refinement results for the examined $Sr_{1-x}La_xFBiS_2$ samples with $^{32}S$ and $^{34}S$ isotopes. In the refinements, a single-phase analysis mode was used, and the La concentration $x$ was fixed as the value determined by EDX.

| Label | #32-1 | #32-2 | #34-1 | #34-2 |
|---|---|---|---|---|
| $x$ (EDX) | 0.387(8) | 0.361(12) | 0.368(8) | 0.361(8) |
| Space group | Tetragonal $P4/nmm$ (#194) | | | |
| $a$ (Å) | 4.084 (3) | 4.084(4) | 4.084(3) | 4.083(4) |
| $c$ (Å) | 13.352(11) | 13.353(13) | 13.366(11) | 13.345(12) |
| $z$ (Sr,La) | 0.1108(2) | 0.1109(2) | 0.1141(3) | 0.1133(3) |
| $z$ (Bi) | 0.62368(14) | 0.6226(2) | 0.6256(2) | 0.6245(2) |
| $z$ (S1) | 0.3703(10) | 0.3780(9) | 0.3638(12) | 0.3692(12) |
| $z$ (S2) | 0.8165(6) | 0.8112(6) | 0.8192(7) | 0.8156(7) |
| $R_{wp}$ (%) | 14.1 | 13.0 | 15.9 | 13.1 |

[Atomic coordinates]
(Sr,La): (0, 0.5, $z$)
F: (0, 0, 0)
Bi: (0, 0.5, $z$)
S1 (in-plane site): (0, 0.5, $z$)
S2 (out-of-plane site): (0, 0.5, $z$)

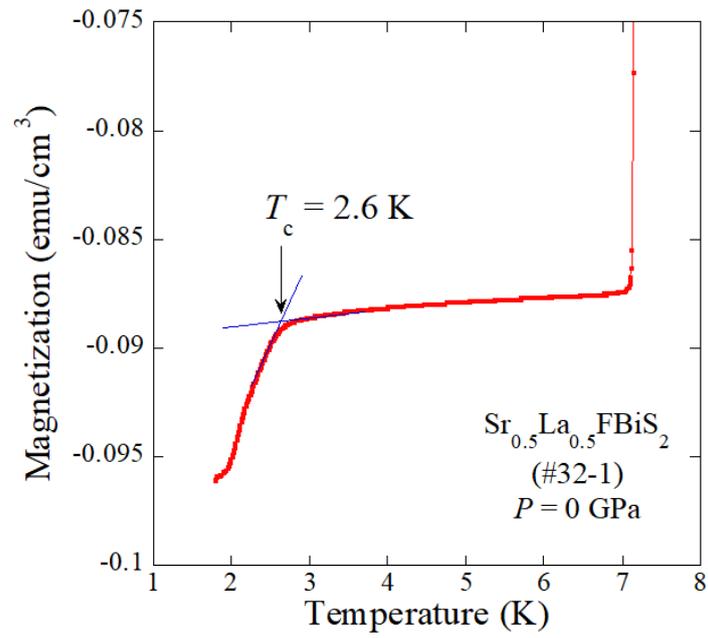

**Fig. S1.** Estimation of $T_c$ for the low-$P$ phase from the temperature dependence of magnetization for #32-1 at ambient pressure.

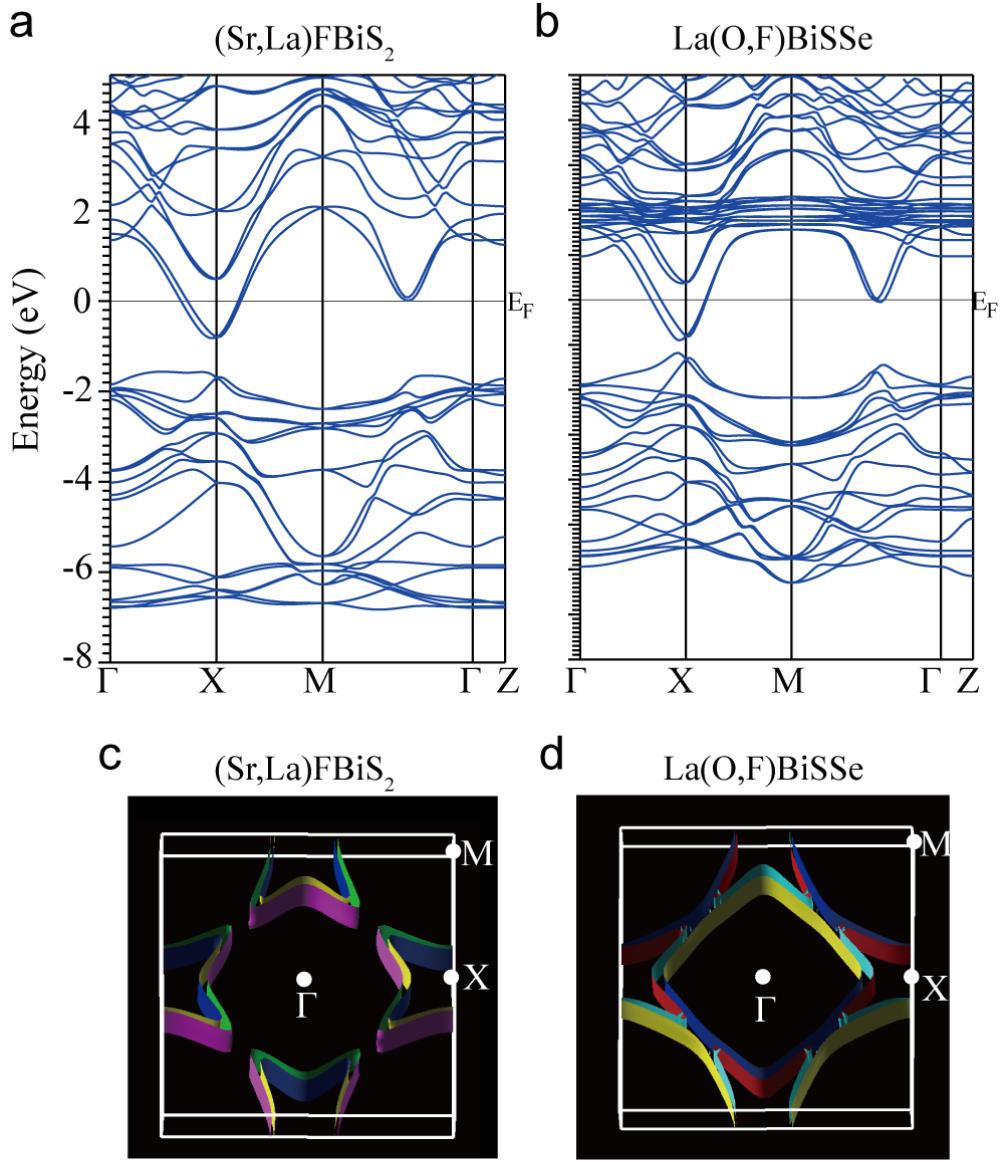

**Fig. S2. Calculated band structure for Sr$_{0.6}$La$_{0.4}$FBiS$_2$ (tetragonal) and LaO$_{0.6}$F$_{0.4}$BiSSe (tetragonal).** (a, c) Calculated electronic band structure and Fermi surface for Sr$_{0.6}$La$_{0.4}$FBiS$_2$. The band calculations were performed from a (Ba,La)FBiS$_2$ model with a lattice structure of Sr$_{1-x}$La$_x$FBiS$_2$ with $x = 0.4$. (b, d) Calculated electronic band structure and Fermi surface for LaO$_{1-x}$F$_x$BiSSe with $x = 0.4$.

**Methods for band calculations**

First-principles band calculations for $Sr_{1-x}La_xFBiS_2$ and $LaO_{1-x}F_xBiSSe$ were performed using the WIEN2k package[1,2]. We used a virtual crystal approximation to simulate partial substitution. Because of technical reasons concerning the virtual crystal approximation, we used Ba instead of Sr for (Sr,La)FBiS$_2$. In the WIEN2k package, the virtual crystal approximation can be performed for elements adjacent to each other in the periodic table. We calculated the electronic band structure of (Sr,La)FBiS$_2$ using the VASP package[3,4], assuming virtual crystal approximation. We have confirmed that the electronic band structure is not strongly affected by the replacement of Sr by Ba in the calculations. The electronic band structures of $Ba_{0.6}La_{0.4}FBiS_2$ and $LaO_{0.6}F_{0.4}BiSSe$ were obtained by adopting the experimental lattice constants of $Sr_{1-x}La_xFBiS_2$ (#32-2: parameters shown in Table S1) and $LaO_{0.6}F_{0.4}BiSSe$[5], respectively. We used $RK_{max} = 7$ and a 18×18×5 $k$-mesh for self-consistent calculations, and adopted the Perdew-Burke-Ernzerhof exchange-correlation functional[6] including the spin-orbit coupling.

1. Blaha, P. *et al.* WIEN2k, An Augmented Plane Wave + Local Orbitals Program for Calculating Crystal Properties (Karlheinz Schwarz, Techn. Universität Wien, Austria), 2018. ISBN 3-9501031-1-2.
2. Blaha, P. *et al.* WIEN2k: An APW+lo program for calculating the properties of solids. *J. Chem. Phys.* 152, 074101 (2020).
3. Kresse, G. & Furthmüller, J. Efficient iterative schemes for ab initio total-energy calculations using a plane-wave basis set. *Phys. Rev. B* 54, 11169 (1996).
4. Kresse, G. & Joubert, D. From ultrasoft pseudopotentials to the projector augmented-wave method. *Phys. Rev. B* 59, 1758 (1999).
5. Hoshi, K., Goto, Y., & Mizuguchi, Y. Selenium isotope effect in the layered bismuth chalcogenide superconductor $LaO_{0.6}F_{0.4}Bi(S,Se)_2$. *Phys. Rev. B* 97, 094509(1-5) (2018).
6. Perdew. J. P., K. Burke & Ernzerhof, M. Generalized Gradient Approximation Made Simple. *Phys. Rev. Lett.* 77, 3865–3868 (1996).